\documentclass[%
reprint,
 amsmath,amssymb,
 aps,
]{revtex4-2}

\usepackage[dvipsnames]{xcolor}
\usepackage{lipsum}
\usepackage{graphicx}

\usepackage{gensymb}
\usepackage{amssymb}
\usepackage{mathtools}
\usepackage{xspace}
\usepackage{upgreek}
\usepackage{hyperref}

\usepackage[]{lineno}
\newcommand{\po}{$^{210}$Po\xspace}
\newcommand{\bi}{$^{210}$Bi\xspace}
\newcommand{\bipo}{$^{214}$Bi-$^{214}$Po\xspace}

\newcommand{\be}{$^{7}$Be\xspace}

\newcommand{\pep}{$pep$\xspace}
\newcommand{\pp}{$pp$\xspace}
\newcommand{\cpd}{cpd/100\,t\xspace}
\newcommand{\ab}{$\alpha/\beta$\xspace}

\begin{document}

\title{Novel techniques for \ab pulse shape discrimination in Borexino}
\author{D.~Basilico\textsuperscript{1}}
\author{G.~Bellini\textsuperscript{1}}
\author{J.~Benziger\textsuperscript{2}}
\author{R.~Biondi\textsuperscript{3,a}}
\author{B.~Caccianiga\textsuperscript{1}}
\author{F.~Calaprice\textsuperscript{4}}
\author{A.~Caminata\textsuperscript{5}}
\author{A.~Chepurnov\textsuperscript{6}}
\author{D.~D'Angelo\textsuperscript{1}}
\author{A.~Derbin\textsuperscript{7}}
\author{A.~Di Giacintov\textsuperscript{3}}
\author{V.~Di Marcello\textsuperscript{3}}
\author{X.F.~Ding\textsuperscript{4,b}}
\author{A.~Di Ludovico\textsuperscript{4,c}} 
\author{L.~Di Noto\textsuperscript{5}}
\author{I.~Drachnev\textsuperscript{7}}
\author{D.~Franco\textsuperscript{8}}
\author{C.~Galbiati\textsuperscript{4,9}}
\author{C.~Ghiano\textsuperscript{3}}
\author{M.~Giammarchi\textsuperscript{1}}
\author{A.~Goretti\textsuperscript{4}}
\author{M.~Gromov\textsuperscript{6}}
\author{D.~Guffanti\textsuperscript{10,d}}
\author{Aldo~Ianni\textsuperscript{3}}
\author{Andrea~Ianni\textsuperscript{4}}
\author{A.~Jany\textsuperscript{11}}
\author{V.~Kobychev\textsuperscript{12}}
\author{G.~Korga\textsuperscript{13,14}}
\author{S.~Kumaran\textsuperscript{15,16,3}}
\author{M.~Laubenstein\textsuperscript{3}}
\author{E.~Litvinovich\textsuperscript{17,18}}
\author{P.~Lombardi\textsuperscript{1}}
\author{I.~Lomskaya\textsuperscript{7}}
\author{L.~Ludhova\textsuperscript{15,16}}
\author{I.~Machulin\textsuperscript{17,18}}
\author{J.~Martyn\textsuperscript{10}}
\author{E.~Meroni\textsuperscript{1}}
\author{L.~Miramonti\textsuperscript{1}}
\author{M.~Misiaszek\textsuperscript{11}}
\author{V.~Muratova\textsuperscript{7}}
\author{R.~Nugmanov\textsuperscript{17}}
\author{L.~Oberauer\textsuperscript{19}}
\author{V.~Orekhov\textsuperscript{10}}
\author{F.~Ortica\textsuperscript{20}}
\author{M.~Pallavicini\textsuperscript{5}}
\author{L.~Pelicci\textsuperscript{15,16}}
\author{\"O.~Penek\textsuperscript{15,f}}
\author{L.~Pietrofaccia\textsuperscript{4,c}}
\author{N.~Pilipenko\textsuperscript{7}}
\author{A.~Pocar\textsuperscript{21}}
\author{G.~Raikov\textsuperscript{17}}
\author{M.T.~Ranalli\textsuperscript{3}}
\author{G.~Ranucci\textsuperscript{1}}
\author{A.~Razeto\textsuperscript{3}}
\author{A.~Re\textsuperscript{1}}
\author{N.~Rossi\textsuperscript{3}}
\author{S.~Sch\"onert\textsuperscript{19}}
\author{D.~Semenov\textsuperscript{7}}
\author{G.~Settanta\textsuperscript{15,g}}
\author{M.~Skorokhvatov\textsuperscript{17,18}}
\author{A.~Singhal\textsuperscript{15,16}}
\author{O.~Smirnov\textsuperscript{22}}
\author{A.~Sotnikov\textsuperscript{22}}
\author{R.~Tartaglia\textsuperscript{3}}
\author{G.~Testera\textsuperscript{5}}
\author{E.~Unzhakov\textsuperscript{7}}
\author{A.~Vishneva\textsuperscript{22}}
\author{R.B.~Vogelaar\textsuperscript{23}}
\author{F.~von~Feilitzsch\textsuperscript{19}}
\author{M.~Wojcik\textsuperscript{11}}
\author{M.~Wurm\textsuperscript{10}}
\author{S.~Zavatarelli\textsuperscript{5}}
\author{K.~Zuber\textsuperscript{24}}
\author{G.~Zuzel\textsuperscript{11}}
% Affiliation
\affiliation{}
\affiliation{\textsuperscript{1}Dipartimento di Fisica, Universit\`a degli Studi e INFN, 20133 Milano, Italy}
\affiliation{\textsuperscript{2}Chemical Engineering Department, Princeton University, Princeton, NJ 08544, USA}
\affiliation{\textsuperscript{3}INFN Laboratori Nazionali del Gran Sasso, 67010 Assergi (AQ), Italy}
\affiliation{\textsuperscript{4}Physics Department, Princeton University, Princeton, NJ 08544, USA}
\affiliation{\textsuperscript{5}Dipartimento di Fisica, Universit\`a degli Studi e INFN, 16146 Genova, Italy}
\affiliation{\textsuperscript{6}Lomonosov Moscow State University Skobeltsyn Institute of Nuclear Physics, 119234 Moscow, Russia}
\affiliation{\textsuperscript{7}St. Petersburg Nuclear Physics Institute NRC Kurchatov Institute, 188350 Gatchina, Russia}
\affiliation{\textsuperscript{8}AstroParticule et Cosmologie, Universit\'e Paris Diderot, CNRS/IN2P3, CEA/IRFU, Observatoire de Paris, Sorbonne Paris Cit\'e, 75205 Paris Cedex 13, France}
\affiliation{\textsuperscript{9}Gran Sasso Science Institute, 67100 L'Aquila, Italy}
\affiliation{\textsuperscript{10}Institute of Physics and Excellence Cluster PRISMA+, Johannes Gutenberg-Universit\"at Mainz, 55099 Mainz, Germany}
\affiliation{\textsuperscript{11}M.~Smoluchowski Institute of Physics, Jagiellonian University, 30348 Krakow, Poland}
\affiliation{\textsuperscript{12}Institute for Nuclear Research of NASU, 03028 Kyiv, Ukraine}
\affiliation{\textsuperscript{13}Department of Physics, Royal Holloway University of London, Egham, Surrey,TW20 0EX, UK}
\affiliation{\textsuperscript{14}Institute of Nuclear Research (Atomki), Debrecen, Hungary}
\affiliation{\textsuperscript{15}Institut f\"ur Kernphysik, Forschungszentrum J\"ulich, 52425 J\"ulich, Germany}
\affiliation{\textsuperscript{16}III. Physikalisches Institut B, RWTH Aachen University, 52062 Aachen, Germany}
\affiliation{\textsuperscript{17}National Research Centre Kurchatov Institute, 123182 Moscow, Russia}
\affiliation{\textsuperscript{18}National Research Nuclear University MEPhI (Moscow Engineering Physics Institute), 115409 Moscow, Russia}
\affiliation{\textsuperscript{19}Physik-Department, Technische Universit\"at  M\"unchen, 85748 Garching, Germany}
\affiliation{\textsuperscript{20}Dipartimento di Chimica, Biologia e Biotecnologie, Universit\`a degli Studi e INFN, 06123 Perugia, Italy}
\affiliation{\textsuperscript{21}Amherst Center for Fundamental Interactions and Physics Department, University of Massachusetts, Amherst, MA 01003, USA}
\affiliation{\textsuperscript{22}Joint Institute for Nuclear Research, 141980 Dubna, Russia}
\affiliation{\textsuperscript{23}Physics Department, Virginia Polytechnic Institute and State University, Blacksburg, VA 24061, USA}
\affiliation{\textsuperscript{24}Department of Physics, Technische Universit\"at Dresden, 01062 Dresden, Germany}

% Present Positions
\affiliation{\textsuperscript{a}Present address: Max-Planck-Institut für Kernphysik, 69117 Heidelberg, Germany}
\affiliation{\textsuperscript{b}Present address: IHEP Institute of High Energy Physics, 100049 Beijing, China}
\affiliation{\textsuperscript{c}Present address: INFN Laboratori Nazionali del Gran Sasso, 67010 Assergi (AQ), Italy}
\affiliation{\textsuperscript{d}Present address: Dipartimento di Fisica, Università degli Studi e INFN Milano-Bicocca, 20126 Milano, Italy}
\affiliation{\textsuperscript{e}Present address: Department of Physics and Astronomy, University of California, Irvine, California, USA}
\affiliation{\textsuperscript{f}Present address: GSI Helmholtzzentrum für Schwerionenforschung GmbH, 64291 Darmstadt, Germany}
\affiliation{\textsuperscript{g}Present address: Istituto Superiore per la Protezione e la Ricerca Ambientale, 00144 Roma, Italy}

\collaboration{The BOREXINO Collaboration -- email: \texttt{spokesperson-borex@lngs.infn.it} }

\date{\today}

\begin{abstract}
Borexino could efficiently distinguish between $\alpha$ and $\beta$ radiation in its liquid scintillator by the characteristic time profile of their scintillation pulse. This \ab discrimination, first demonstrated at the tonne scale in the Counting Test Facility prototype, was used throughout the lifetime of the experiment between 2007 and 2021. With this method, $\alpha$ events are identified and subtracted from the $\beta$-like solar neutrino events. This is particularly important in liquid scintillator as $\alpha$ scintillation is quenched many-fold. In Borexino, the prominent \po decay peak was a background in the energy range of electrons scattered from \be solar neutrinos. Optimal \ab discrimination was achieved with a \emph{multi-layer perceptron neural network}, which its higher ability to leverage the timing information of the scintillation photons detected by the photomultiplier tubes. 
An event-by-event, high efficiency, stable, and uniform pulse shape discrimination was essential in characterising the spatial distribution of background in the detector. This benefited most Borexino measurements, including solar neutrinos in the \pp chain and the first direct observation of the CNO cycle in the Sun. This paper presents the key milestones in \ab discrimination in Borexino as a term of comparison for current and future large liquid scintillator detectors.

%The Borexino collaboration has consistently employed the \ab statistical subtraction technique since the inception of data collection in 2007. This method was initially exploited in the Counting Test Facility prototype. This strategy serves to effectively mitigate and regulate potential biases arising from the prominent \po peak within the \be neutrino spectrum, as a consequence of the strong $\alpha$-quenching of the scintillator target. Improvements to the \ab discrimination have been achieved through the implementation of a \emph{multi-layer perceptron neural network}. This novel method has shown a better capability in leveraging input features derived from the timing of photoelectron arrivals, captured by the photomultiplier tubes responsible for detecting scintillation light. A consistently high efficiency, stable, and uniform pulse shape discrimination on an event-by-event basis has emerged as a pivotal determinant in accurately characterising the spatial distribution of background contamination. This achievement holds immense significance for numerous analyses conducted within the Borexino project, most notably the precision measurement of solar neutrinos within the \pp chain, as well as the fist  observation of the CNO cycle in the Sun. Within this article, we comprehensively survey the key milestones achieved in unravelling the \ab discrimination capability of Borexino. These accomplishments serve as a reproducible foundation for both existing detectors and future experiments employing analogous liquid scintillator technologies.
\end{abstract}

%\keywords{keywords{machine learning, multilayer perceptron, artificial neural network, pulse shape discrimination, solar neutrinos, scintillation detectors}

\maketitle

\section*{Introduction}

For as long as it operated, Borexino was the only detector capable of measuring %the 
solar neutrino interactions (position and energy) on an event-by-event basis %interaction rate 
with a threshold %of 
$\gtrsim 150$ keV, \emph{i.e.}, down to the $^{14}$C $\beta$-spectrum end-point. %and of reconstructing the position and energy on an event-by-event basis. 
An important feature of Borexino was the possibility to efficiently separate events initiated by recoiling electrons ($\beta$-like events) versus $\alpha$ particles. The former include solar neutrino interactions as well as background from $\beta$ and $\gamma$ decays. This is possible via pulse-shape discrimination (PSD) techniques that exploit the different time profile of the scintillation emission for $\alpha$ and $\beta$-like events (see, \emph{e.g.}, \cite{bib:long}).
%Furthermore, thanks to the peculiar time distribution of the scintillation light, spanning over about 1.5 $\upmu$s, and the event topology, it was possible to exploit a pulse-shape discrimination (PSD) to perform a high efficiency separation of the electron recoils induced by solar neutrino interaction and of the $\beta-\gamma$ contamination (hereafter, $\beta$-like events) from the $\alpha$ background (for review, see \emph{e.g.} \cite{bib:long}).  
The so-called \ab discrimination played an important role in solar neutrino measurements throughout the Borexino data taking between 2007 and 2021. It is worth noting that Borexino also achieved $\beta^-/\beta^+$ separation via PSD, as reported in~\cite{bib:pep} and~\cite{bib:global}; the latter topic is, however, outside the scope of the present article. 

PSD for \ab separation was first studied within the Borexino program with the 4-tonne ``Counting Test Facility'' (CTF) prototype \cite{bib:ctf-1, bib:ctf-2}. The original method is based on the \emph{Gatti} parameter \cite{bib:gatti}  and enabled a statistical subtraction of $\alpha$ background, especially from \po, from the measured energy spectrum. Monochromatic, 5.3 MeV \po $alpha$s ($Q_{\alpha}$=5407 keV) appeared in the Borexino liquid scintillator as a peak at $\sim$500 keV of electron-equivalent energy due to a greater than ten-fold quenching of the scintillation for these highly ionizing tracks~\cite{bib:ctf-2}. At the start of the Borexino data taking, the \po rate was $\sim8000$ counts per day per 100 tonnes (hereafter, \cpd). 
%to remove and to control any possible bias due to the presence of a high rate \po monochromatic $\alpha$  line, peaked at $\sim$500 keV and with activity of the order of 8000 counts per day per 100 tonnes (hereafter, \cpd) at the beginning of the data acquisition in 2007. The actual Q-value of the \po decay is 5407 keV, however its equivalent energy on the electron-like radiation scale is reduced by a factor $\sim$13 by a strong energy dependent scintillation quenching~\cite{bib:ctf-2},  
Quenching was also observed for other $\alpha$ particles. These include those from the thoron ($^{220}$Rn) and radon ($^{222}$Rn) decay chains which are handily identified using their time coincidence, \emph{e.g.}, $^{212}$Po (8954 keV), $^{214}$Po (7833 keV), and $^{218}$Po (6114 keV).
%, easily identifiable through the fast decay coincidences part of the decay chain of the thoron ($^{220}$Rn) and radon ($^{222}$Rn) decays, respectively.

Because of its quenching, the \po peak falls within the \be solar neutrino Compton-like energy spectrum, which presents a characteristic shoulder at 662 keV. %above the detector energy threshold. 
Although in this case the \be shoulder appears at higher energy than the \po peak, making it possible for the multi-parameter spectral fit to clearly identify these two separate components, the Borexino analysis was performed both with and without bin-by-bin statistical \ab subtraction of the \po peak to ensure that there was no subtle bias due to the presence of the $\alpha$ background. This was particularly true in Phase-I of the experiment, when the \po activity was more than two orders of magnitude greater than the \be event rate ($\sim$50 \cpd over the entire energy range).
In each bin, we assumed the Gatti parameter to be normally distributed with a mean value linearly dependent on energy~\cite{bib:be7-1, bib:be7-2, bib:be7-3}. 

%The prominent \be shoulder is anyway visible above the \po peak and makes the spectral fit disentangle easily the signal from the background for this particular neutrino species. Anyway, the \po activity is much higher as compared with the \be interaction rate (only $\sim$50 \cpd on the full energy interval), especially in Phase-I. In order to control any possible bias due to overwhelming \po peak, the analysis is performed either fitting the spectrum with or without the \po peak. In the latter, the $\alpha$ events are ``statistically subtracted'' in each histogram bin, according to the distribution of the Gatti parameter assumed to be Gaussian with the mean value linearly dependent on the energy \cite{bib:be7-1, bib:be7-2, bib:be7-3}.

As data-taking progressed, 
%After about four years from the beginning of data taking in mid-2007, 
the \po naturally decayed with a lifetime $\tau=199.6$ d.
The reduction of this background  was, however, counterbalanced by a progressive degradation of the energy resolution of the detector
%with a consequent mitigation of any possible systematic bias of the $\alpha$ line in the subsequent Borexino phases. However, the resolution degradation 
due to the loss of photomultiplier tubes (PMTs). 
The count of working PMTs decreased for $\sim$2000 units in mid 2007 to $\sim$1000 units at the end of 2021.
This effect and the need for a more uniform, stable, and higher efficiency \ab discrimination
%with high efficiency, 
for the study of CNO solar neutrinos suggested exploring 
%for the study of the \po contamination for to the CNO analysis, has pushed the Collaboration to improve the \ab selection, exploiting 
novel techniques based on neural networks already extensively employed in particle physics.
We pursued 
%This goal was satisfactorily achieved by exploiting 
neural networks based on \emph{multi-layer perceptron} (MLP). 
The subject of this paper is the description of the MLP input parameters, structure, and training strategy given the Borexino scintillator properties and layout. It also presents studies of the network's efficiency for \ab discrimination. 
%To reach the desired performances, an intense investigation was needed for the identification of the input parameters and for the optimisation of the structure of the neural network. The main topic of this article is the description of MLP training strategy and the efficiency studies, adapted to the scintillation properties and the geometry of the Borexino detector.

In Sec.~\ref{sec:bx}, the main characteristics of the Borexino detector and its main physics results relevant to this article are briefly reviewed. In Sec.~\ref{sec:gatti}, the \ab PSD in Borexino using the Gatti parameter is presented. In Sec.~\ref{sec:mlp}, the implementation strategy of the MLP on the Borexino scintillator time profile is described along with 
an evaluation of its performance and efficiency. Finally, in Sec.~\ref{sec:cno}, the impact of \ab MLP discrimination on the CNO solar neutrino analysis and on other Borexino results over its 14 years lifetime is discussed.

\section{The Borexino Experiment} \label{sec:bx}
\begin{table*}[t!]
\begin{center}
\begin{tabular}{lll}
\hline \hline 
Species & Rate [\cpd] & Flux [cm$^{-2}$ s$^{-1}$ ] \\ \hline
{\it pp}     & $(134 \pm 10)^{+6}_{-10}  $ &  $(6.1 \pm 0.5)_{-0.5}^{+0.3} \times 10^{10}$   \\
$^7$Be & $(48.3 \pm 1.1)^{+0.4}_{-0.7}$ &  $(4.99 \pm 0.11)_{-0.08}^{+0.06} \times 10^9$ \\
{\it pep}  (HZ)   & $(2.7 \pm 0.4)^{+0.1}_{-0.2} $  & $(1.3 \pm 0.3)_{0.1}^{+0.1} \times 10^8$ \\ 
$^8$B($> 3$ MeV)& $ 0.223_{-0.022}^{+0.021}$ & $ 5.68_{-0.44}^{+0.42} \times 10^6$ \\
{\it hep} & $< 0.002$ (90\% CL) & $<1.8 \times 10^5$ (90\% CL)   \\
CNO    & $6.7_{-0.8}^{+1.2}$    &  $6.7_{-0.8}^{+1.2} \times 10^8$ 
\\ \hline \hline
\end{tabular}
\end{center}
\caption{Solar neutrino interaction rates in Borexino and extrapolated solar neutrino fluxes for the different components of the \pp chain and CNO cycle. Rates are reported in \cpd, while fluxes are reported in cm$^{-2}$s$^{-1}$. N.B.: HZ stands for \emph{high metallicity} assumption.}
\end{table*} \label{tab:solars}

Borexino was located in the Hall C of Laboratori Nazionali Gran Sasso (LNGS) of the Italian Institute of Nuclear Physics (INFN) \cite{bib:lngs}. The detector had been taking data from mid-2007 to the end of 2021, and is currently under decommissioning.
The detector is made of concentric layers of increasing radiopurity (see for details \emph{e.g.} \cite{bib:tech}): the innermost core, called Inner Vessel (IV), consists of about 280 tons of liquid scintillator (\emph{pseudocumene} mixed with 1.5 g/l of PPO as scintillating solute) contained inside an ultra-pure nylon vessel with a thickness of 125 $\upmu$m and a radius of 4.25 m. A Stainless Steel Sphere (SSS), filled up with the remaining $1000$ m$^3$ of buffer liquid (\emph{pseudocumene} mixed with DMP quencher) is instrumented with more than 2000 PMTs for detecting the scintillation light inside the IV. Finally, the SSS is immersed in an about 2000 m$^3$ Water Tank (WT), acting as Cerenkov veto, equipped with 200 PMTs. Using results from the study of internal residual contaminations and from the 2010 calibration, it results that the detector is capable of determining the event position with an accuracy of $\sim 10$ cm (at 1 MeV)  and the event energy with a resolution following approximately the relation $\sigma(E) / E \simeq 5\%/\sqrt{E/[MeV]}$. 

The Borexino data set is divided, according to the internal conventional subdivision of the experimental program, in three different phases: Phase-I, from May 2007 to May 2010, ended with the calibration campaign, in which the first measurement of the \be solar neutrino interaction rate  \cite{bib:be7-1, bib:be7-2, bib:be7-3} and the first evidence of the \pep \cite{bib:pep} were performed; Phase-II, from December 2011 to May 2016, started after an intense purification campaign with unprecedented reduction of the scintillator radioactive contaminants, in which a 10\% first spectroscopic observation of the \pp neutrinos \cite{bib:pp} was published, and later updated in the solar neutrino comprehensive analysis of all \pp chain  neutrino fluxes \cite{bib:global, bib:nusol, bib:b8}; finally, Phase-III, from July 2016 to October 2021, after the thermal stabilisation program, in which the first detection of the CNO neutrinos \cite{bib:cno} and its subsequent improvements \cite{bib:PRLcno, bib:CID} were achieved.
The most important solar neutrino results in terms of interaction rate and corresponding fluxes are summarised in Tab.~\ref{tab:solars}.
Thanks to its unprecedented radio-purity, Borexino has also set a lot of limits on rare processes \cite{bib:elec, bib:nsi, bib:astro-nu, bib:magnet, bib:sterile} and performed other neutrino physics studies, as \emph{e.g.} geo-neutrino detection (for review, see \emph{e.g.} \cite{bib:geonu}). As it will be highlighted in the Sec. \ref{sec:cno}, all these important results are strongly dependent on the \ab PSD optimisation. In the following Section, the \ab discrimination problem is introduced, starting from the first method exploited by the Collaboration, based on the Gatti parameter. 

\section{$\alpha/\beta$ discrimination: the \emph{Gatti} parameter} 
\label{sec:gatti}

\begin{figure}[t!]
\begin{center}
\centering{\includegraphics[width=1.0\columnwidth]{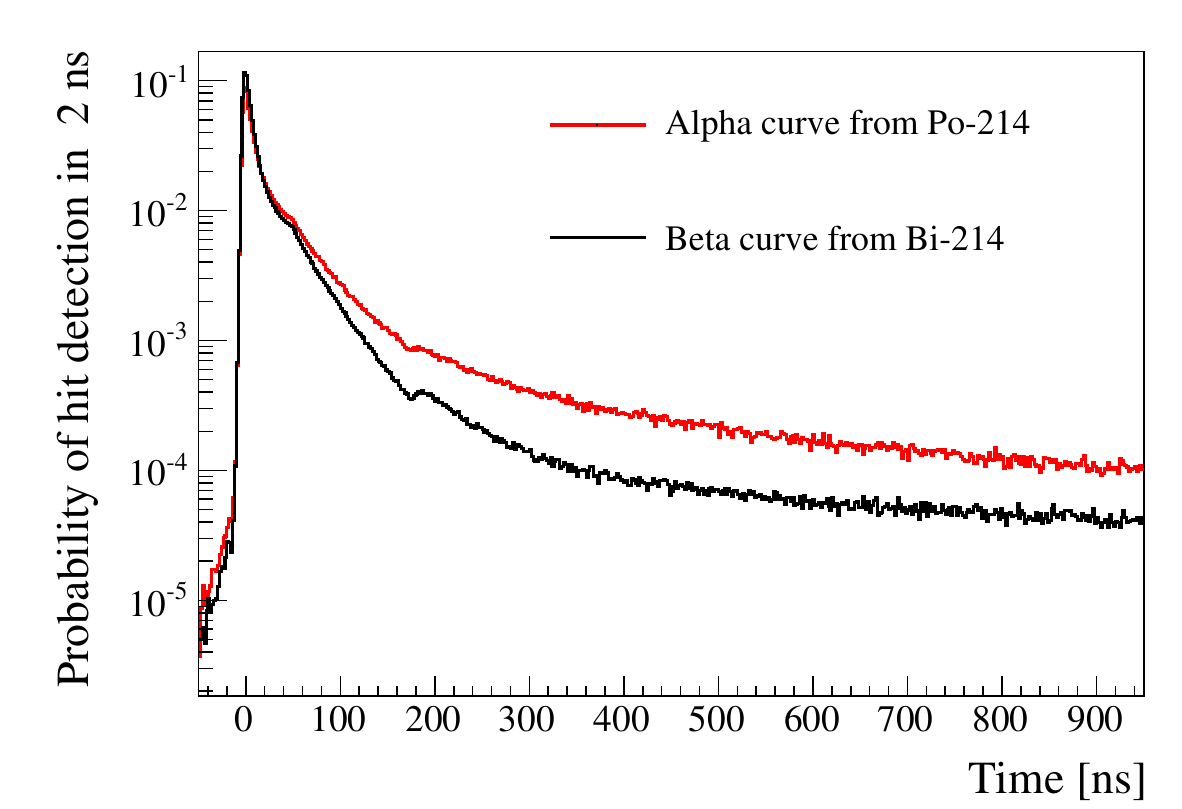}}
\caption{The reference $P_{\alpha}(t)$ (red) and $ P_{\beta}(t)$ (black) pulse shapes obtained by tagging the $^{222}$Rn-correlated $^{214}$Bi-$^{214}$Po coincidences. The dip at 180\,ns is due to the dead time on every individual electronic channel applied after each detected hit.  The small knee around 60\,ns is due to the reflected light on the SSS surface and on the PMTs photo-cathodes.} 
\label{fig:abPulse}
\end{center}
\end{figure}
\begin{figure} [h!]
\begin{center}
\centering{\includegraphics[width=1.0\columnwidth]{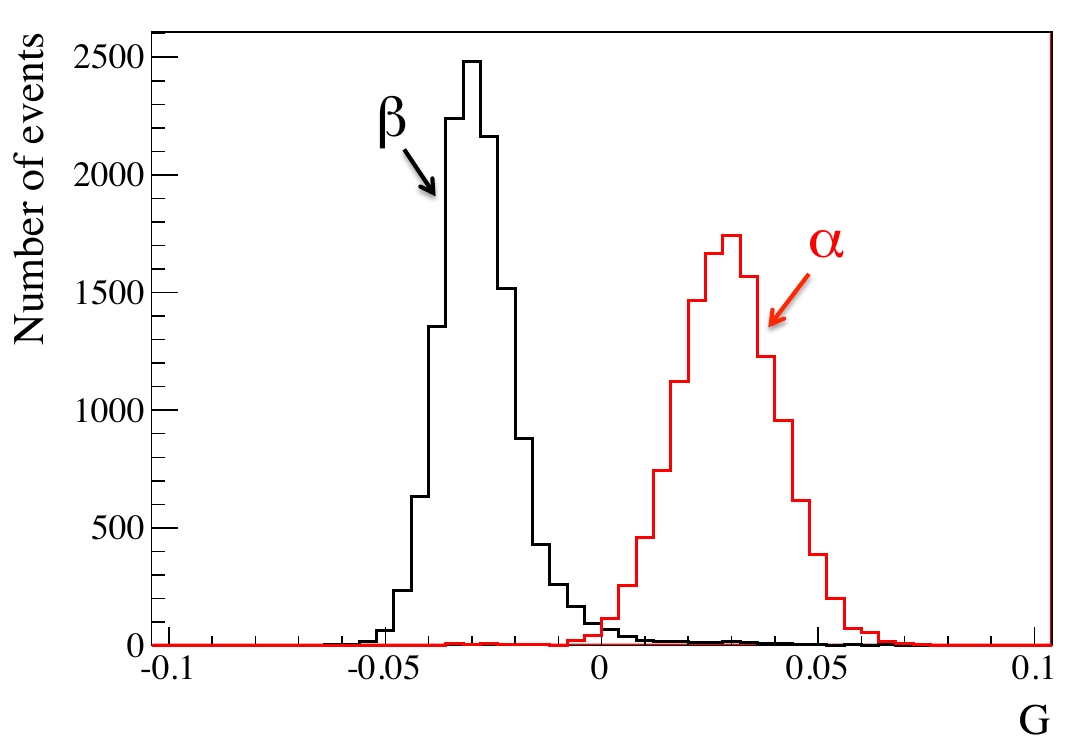}}
\caption{The distribution  of $G_\alpha$ (red) and $G_\beta$ (black) (see Eq. \ref{eq:GattiDef}) for events obtained by tagging the radon correlated $^{214}$Bi--$^{214}$Po coincidences.}
\label{fig:Gattiab}
\end{center}
\end{figure}

The $\alpha$/$\beta$ discrimination in Borexino is possible thanks to
the sizeable difference between the time distributions of the scintillation light (pulse shape) for $\alpha$ and $\beta$-like events (see Fig.~\ref{fig:abPulse}).
For each event meeting the threshold condition of a few tens of photo-electrons (PE) in 100 ns ($\sim60$ keV), the arrival times of PEs on each PMTs are recorded.  As described in detail in~\cite{bib:long}, for each detected PE, the arrival time and the charge are measured by an analogue and digital electronics chain.  When a trigger occurs, the time and the charge of each PMT, that has detected at least one photoelectron in a time gate of 7.2\,$\upmu$s, is recorded.  The time is measured by a time-to-digital converter (TDC) with a resolution of about 0.5\,ns, while the charge (after integration and pulse shaping) is measured by an 8 bit analogue-to-digital converter (ADC). The time resolution is smaller than the intrinsic PMT time jitter of about 1.1\,ns.

For each event, time, charge, and position are reconstructed by the offline software. The code identifies in the recorded time window of 16 $\upmu$s a group of time correlated hits, called ``cluster''. Each event is generally made of a single cluster, but in the case of fast coincidences, like $^{214}$Bi-$^{214}$Po and $^{85}$Kr-$^{85m}$Rb close decays, or in case of accidental pile-ups (often occurring with very frequent $^{14}$C events), more than one cluster could be identified as superposition of two of more different events.
The position of events is determined via a photon time-of-flight maximum likelihood method with probability density functions (PDF) based on experimental data and Monte Carlo simulations, resulting in an uncertainty of 10 cm for each of the three Cartesian spatial coordinates. The spatial resolution is expected to scale naively as $1/\sqrt{NPE}$ where NPE is the number of detected photoelectrons~\cite{bib:galby}. 
For the most important analyses in Borexino, the fundamental event selection is based on the following criteria: internal only trigger (no muon veto coincidence), event time  2\,ms off a preceding muon event, single cluster in the acquisition window and position reconstructed in $r\lesssim 3$\,m. These cuts guarantee that the selected event is a neutrino-like candidate, i.e. an event occurred in the innermost part of the IV ($\lesssim$100 t) and far enough from the external background coming from the SSS and from the IV structures. 

After the application of the selection criteria listed above, the typical Borexino spectrum shows a prominent $^{210}$Po $\alpha$ peak at about  500  keV, that falls inside the $^7$Be energy window, see \emph{e.g.}~\cite{bib:be7-2}.  
At the beginning of Phase-I the \po activity was of order $10^4$ \cpd. At the beginning of Phase-II, more than 4 years later, the activity went down by one order of magnitude to $\sim10^3$ \cpd, 
a bit more than expected because a little amount of \po was reintroduced by the water extraction campaign.
Finally in Phase-III, after more than 4 years and thanks to the thermal insulation campaign,  which reduced drastically the scintillator convective motions (see Sec. \ref{sec:cno} for further details), the \po activity was significantly lowered by another order of magnitude, namely  $\sim10^2$ \cpd. This allowed one to reach the condition of the CNO measurement via the so-called \bi-\po link \cite{bib:cno}.

An estimation of the \po activity and its possible independent quantification for the Borexino analysis can be done for example by defining a simple parameter called \texttt{tail-to-tot} (\texttt{t2t}), which is defined as the fractional portion of the time distribution of the hits above a given characteristic time $t_0$ with respect to the beginning of the scintillation, namely
\begin{equation} \label{eq:t2t}
    {\rm\ \texttt{t2t}} = \frac{\int\limits_{t_0}^\infty S(t)\mathrm{d}t}{\int\limits_{0}^\infty S(t) \mathrm{d}t},
\end{equation}
where $S(t)$ is the scintillation time distribution. The characteristic time $t_0$ can be optimised by maximising the figure of merit defined as the difference between the \texttt{t2t} populations for $\alpha$ and $\beta$ events.
This sort of parameter works very well for example for separating electron and nuclear recoils in liquid argon scintillation chambers \cite{bib:f90}, where the scintillation light is basically made of a combination of two typical exponential decay times, differing 3 order of magnitude from each other (typically 6 and 1600 ns). This is not the case of the Borexino scintillation, where the time behaviour is more complicated and less specific for different particle types \cite{bib:ranucci}. As a consequence, \texttt{t2t} in Borexino gives a more mild \ab separation rather than a real high efficiency event classification. 

\begin{figure}[!t]
\centering
\includegraphics[width=1.0\columnwidth]{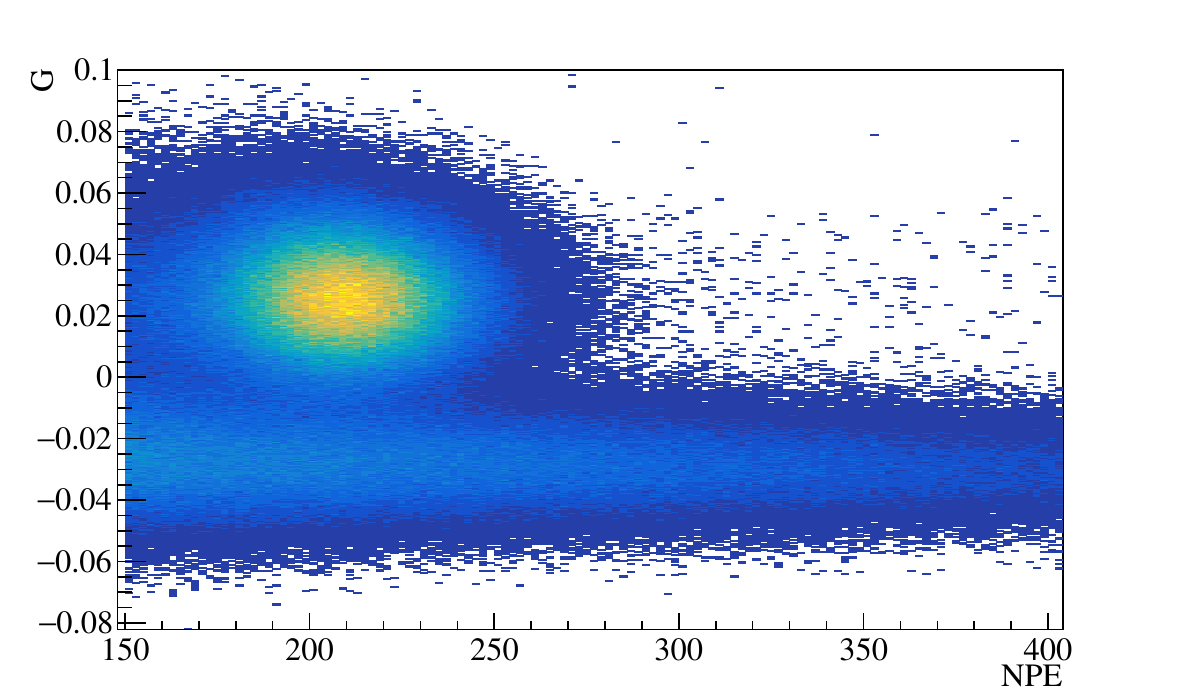}
\caption{Example of \ab separation in the Gatti-Energy(NPE) space (first 300 days of Borexino Phase-II) . The big blob on the top represents the $\alpha$ distribution (\po), while the bottom horizontal belt represents the $\beta$-like component.}
\label{fig:GNPE}
\end{figure}
\begin{figure}[!t]
\centering
\includegraphics[width=1.0\columnwidth]{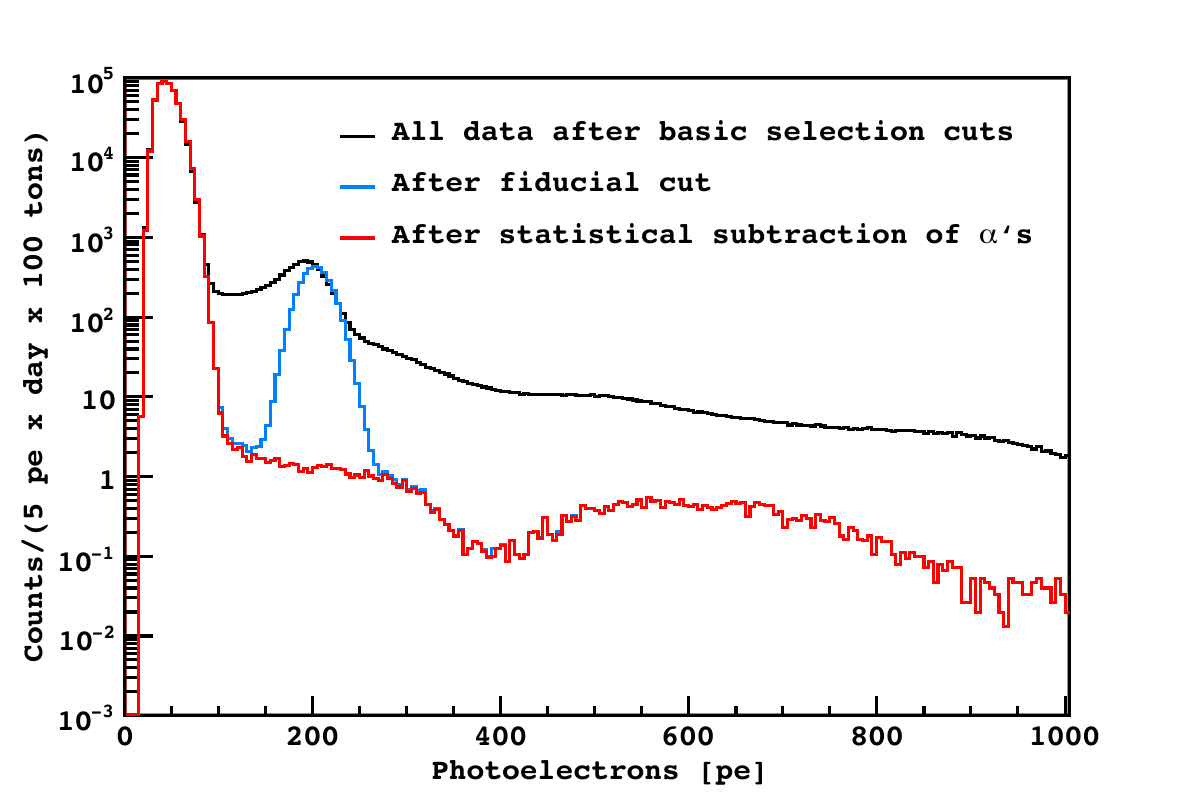}
\caption{The raw NPE charge spectrum after the basic selection criteria (black), after the fiducial volume cut (blue), and after the statistical subtraction of the $\alpha$-emitting contaminants (red).}
\label{fig:spectra}
\end{figure}

A more efficient identification of $\alpha$/$\beta$, instead, can be performed using discriminating procedures like the Gatti optimal filter~\cite{bib:gatti}.
The latter allows one to classify two
types of events with different, but known, time distributions of hits as a function of time.
Their reference shapes $P_\alpha(t)$ and
$P_\beta(t)$ are created by averaging the time distributions of a large sample of events selected independently, without any use of pulse shape variables. A practical way to build the reference shapes is to use the $^{214}$Bi-$^{214}$Po fast coincidence, originating from the $^{222}$Rn events in the scintillator. The $^{214}$Bi-$^{214}$Po coincidences (a few hundreds of $\upmu$s) in Borexino are tagged with a space-time correlation with about 90\% efficiency, basically limited by the trigger threshold for the preceding $\beta$ event. The first of the two events in time provides a pure $\beta$-like sample with events mostly located in the energy interval 1500-3000 keV as superposition of different $\beta$ and $\gamma$ lines, while the second provides a pure $\alpha$ sample with events peaked at about 800 keV and smeared only by the detector resolution. The radon events in the IV (with about one week lifetime) are strictly related to invasive operations on the scintillator (especially at the beginning of Phase-I and during the WE campaign), while are basically absent in quiet periods as  Phase-II and especially Phase-III.

The functions $P_\alpha(t)$ and $P_\beta(t)$ represent the PDFs as a function of time of detecting a PE for events of type $\alpha$ or $\beta$, respectively. 
Let $e(t)$ be the normalised time
distribution of the light for each event. The Gatti parameter $G$ is defined as
\begin{equation}
G = \int e(t) w(t) dt,
\label{eq:GattiDef}
\end{equation}
where $w(t)$ are the weights given by
\begin{equation}
w(t)  = \frac{P_\alpha(t)- P_\beta(t)}{P_\alpha(t)+P_\beta(t)} \cdot
\label{Gattiw}
\end{equation}
The $G$ parameter follows a probability distribution with the mean
value $\langle G_{\alpha, \beta} \rangle$, which depends on the particle type, namely
\begin{equation} \label{eq:amu}
\langle G_{\alpha, \beta} \rangle= \int P_{\alpha, \beta}(t) w(t) dt. 
\end{equation}
In the Borexino scintillation, the Gatti mean values are empirically found to be linearly decreasing with energy. 
Finally, considering the Poissonian statistical fluctuations of the entries in each time bin,  the corresponding variance, following the variance expansion identity,  reads 
\begin{equation} \label{eq:asigma}
\sigma_{G_{\alpha, \beta}}^2 = \int P^2_{\alpha, \beta}(t) w(t)dt - \langle G_{\alpha, \beta} \rangle^2.
\end{equation}

In the real experimental case, the integration in Eqs.~\ref{eq:amu} and~\ref{eq:asigma} are converted into a sum over histograms binned at 1\,ns from zero to about 1.5\,$\upmu$s.
In the scintillator used by Borexino, $\alpha$ pulses are slower and have therefore a longer tail with respect to $\beta$ pulses. This feature represents basically the key for the \ab separation.
Examples of reference shapes $P_{\alpha} (t)$ and $P_{\beta}(t)$, from the $^{214}$Bi-$^{214}$Po tagging of the radon events from Phase-I, are shown
in Fig.~\ref{fig:abPulse}: the dip at 180 ns is due
to the dead time on every individual electronic channel applied after each detected hit. The small knee
around 60 ns is due to the reflected light on the SSS
surface and on the PMTs photo-cathodes. The distributions of the corresponding G parameters ($G_{\alpha}$ and $G_{\beta}$) for events with respect to these reference PDFs  are shown in Fig.~\ref{fig:Gattiab}. The two distributions, resembling Gaussian shapes,  are partially overlapped due to the sizeable $G$ variance. As a consequence, when the number of $\alpha$ events largely exceeds that of the $\beta$'s, a
high efficiency event-by-event \ab selection is anyway limited. In principle, a bin-by-bin statistical separation of the two event populations is possible, whenever the $G_{\alpha, \beta}$ distribution are known either analytically or through a Monte Carlo simulation. 
Since the mean values and the variances of $G_{\alpha, \beta}$ are energy dependent,  their distributions are fitted to
two Gaussian models for each bin in the energy
spectrum of interest, and their value are forcibly constraint around the linear dependence guess.
The integral of the fitted curves represents the relative contribution of each species in each energy bin, 
and the $\alpha$ contribution is subtracted from the total bin content, thus obtaining the $\beta$-like spectrum by statistical subtraction.

\begin{figure}[t]
\begin{center}
\centering{\includegraphics[width=1.0\columnwidth]{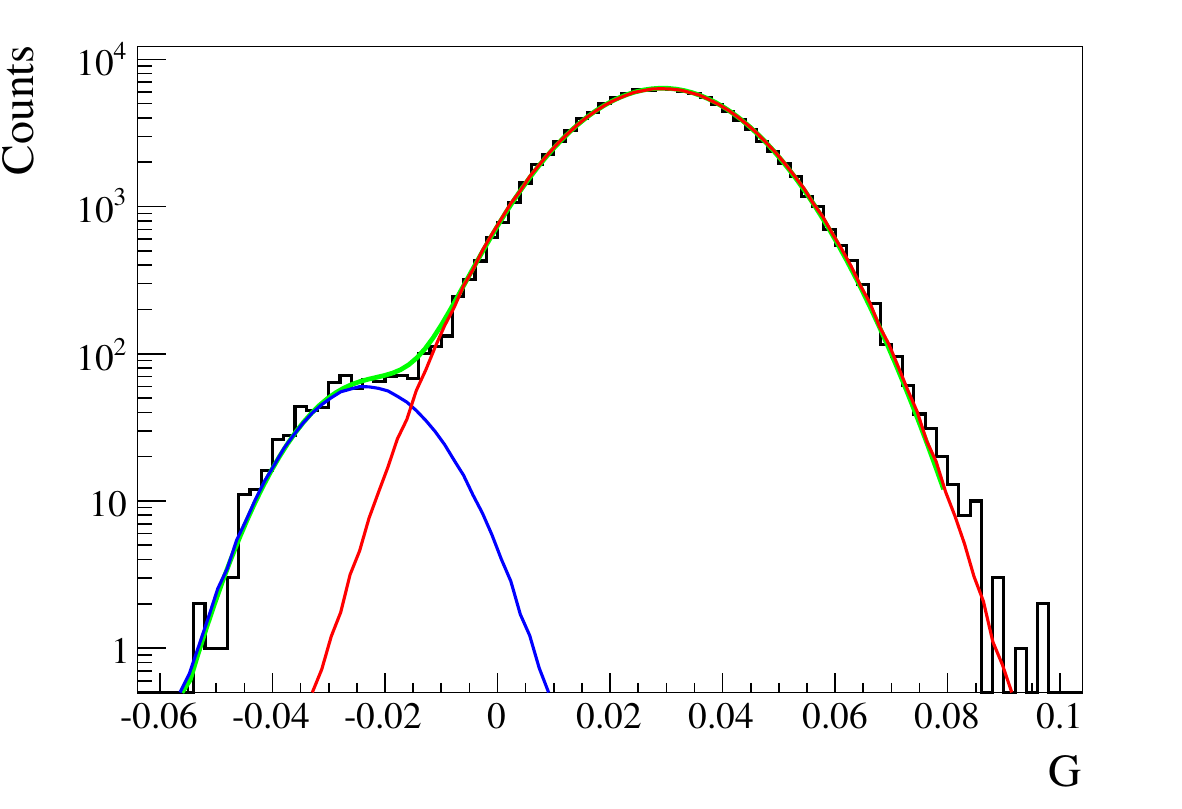}}
\caption{Example of \ab statistical subtraction with
  the analytical curves for events in the energy range 200-205 NPE. The blue and red lines show the individual Gaussian fits to
  the Gatti parameter distributions for the $\beta$ and $\alpha$
  components, respectively, while the green line is the total fit.}
\label{fig:abAnalytical}
\end{center}
\end{figure}

We make the reasonable hypothesis that the underlying distributions are Gaussian.
The fit procedure also provides the error of the estimated particle population to be replaced in the corresponding bin in which the subtraction is performed.
In bins where one
species greatly outnumbers the other, for example in the energy bins in which the $^{210}$Po is peaked, the mean values of the Gaussian parameters are fixed to their predicted values, extrapolated from the energy dependency trend, in order to avoid any possible bias in the subtraction procedure.
Figure~\ref{fig:abAnalytical} shows an example of
the $G_{\alpha \beta}$ parameter in the energy range bin 200-205 NPE 
and its fit to the analytical
model. Furthermore any other possible double Gaussian fit bias, due to the large difference in the two population statistics,  is corrected according toy Monte Carlo simulations with the same population ratio. 

The statistical subtraction can be applied in the full $^7$Be energy window, removing all  $\alpha$ backgrounds coming mostly from 
$^{210}$Po, but also from other $^{222}$Rn $\alpha$'s daughters such as $^{214}$Po and $^{218}$Po leaking the fast coincidence cut. This secondary subdominant contamination is partially affecting Phase-I, but is it completely negligible in Phase-II and Phase-III, thanks to the better background condition achieved after the WE campaign and \po decay.
The error associated to the statistical subtraction is propagated as a systematic uncertainty on the final neutrino interaction rates \cite{bib:be7-1, bib:be7-2, bib:be7-3}. It is worth mentioning that a possible bias due to the presence of the \po peak is not negligible only in Phase-I when the \po activity is much larger than \be neutrino interaction rates. In fact, in Phase-II and in Phase-III the statistical subtraction of the $\alpha$ component is not applied and the \po is simply quantified by the spectral fit, see \emph{e.g.} \cite{bib:global,bib:cno}. 

Figure~\ref{fig:GNPE} shows the Gatti distribution as a function of the event energy in NPE for the first 300 days of Borexino Phase-II. The big blob on the top represents the $\alpha$ distribution, consisting basically of \po events, while the bottom horizontal belt represents the $\beta$-like component (solar neutrinos and background). The Gatti parameter shows a neat separation of the \ab population as a function of the energy. 

Figure~\ref{fig:spectra} shows the implementation of the \ab statistical subtraction in Phase-I: the black curve represents the energy distribution of all events before applying the basic selection criteria. The blue curve represents the event energy distribution after the fiducial volume selection: below 100~NPE  the spectrum is dominated by $^{14}$C decay ($\beta^-$, $Q$=156~keV) ~\cite{bib:14c} and the peak at 200~NPE is dominated by $^{210}$Po~decays.  The red curve is the final spectrum after the statistical subtraction of the $\alpha$ component.  
The prominent feature around 300 NPE includes the Compton-like edge due to \be solar neutrinos. Finally,  the large bump peaked around $\sim$600 NPE is the spectrum of the cosmogenic $^{11}$C ($\beta^+$, $Q$=1.98~MeV, created in situ by cosmic ray-induced showers).

Thanks to its relevant discrimination power, the \ab based on the Gatti parameter has been applied in many Borexino analyses also as {\it soft cut} for pre-selecting events and {\it hard cut} to locate the main $\alpha$ contaminants, as the $^{210}$Po within the fiducial volume, and for understanding the nature of the main backgrounds \cite{bib:long}, \emph{e.g.} in the geo-neutrino analysis \cite{bib:geonu}. In those cases, no statistical subtraction is applied, and the Gatti parameter selects the $\alpha$ population with a given efficiency, depending on the position of the cut itself.

\begin{figure}[!t]
\centering
\includegraphics[width=1.0\columnwidth]{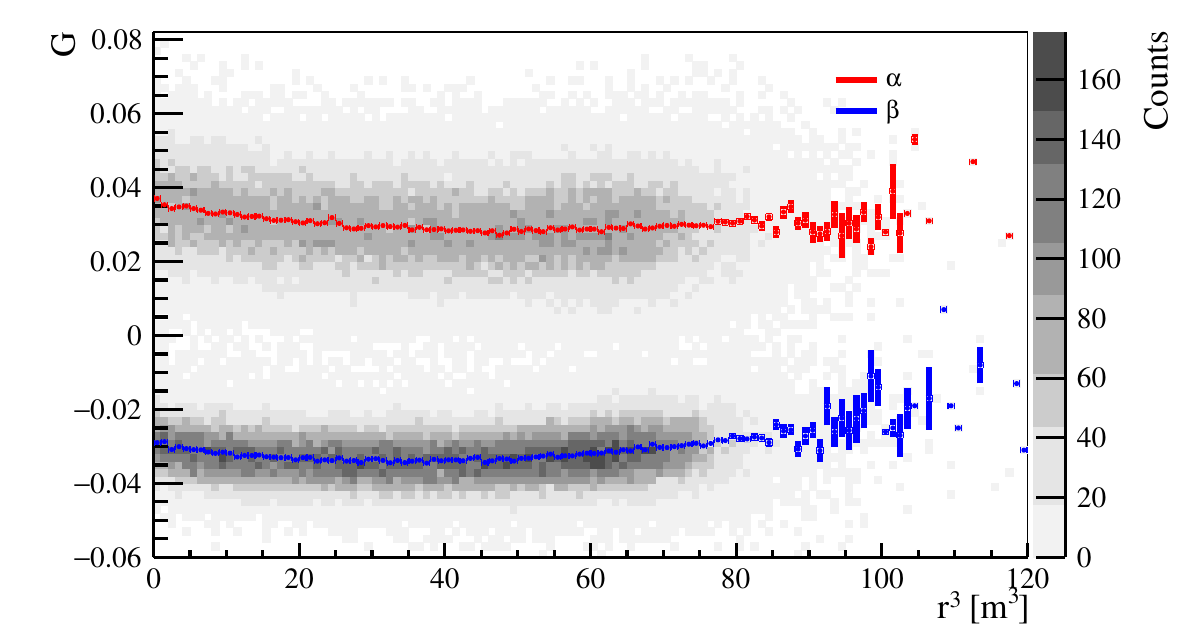}
\caption{Radial ($r^3$) dependence of the Gatti parameter $G$ (grey) for a sample of \bipo events. Red and blue points represent the mean values with their uncertainty for $\alpha$ and $\beta$-like events, respectively.}
\label{fig:gatti-r}
\end{figure}

The optimisation of the Gatti filter, already exploited in the Borexino CTF, played a crucial role in many important Borexino studies. Nevertheless new requirements and some drawbacks pushed the Collaboration to investigate other novel techniques based on neural networks. As it will be described in Sec.~\ref{sec:cno}, the CNO feasibility study had been requiring, since the beginning of Phase-II, a deep understanding of the spacial evolution of the \po contamination. This analysis required, instead of a statistical subtraction, a high efficiency event-by-event selection uniform in space, and easily modelable in energy and time. The PMT loss, and the consequent resolution degradation, is affecting the Gatti parameter distributions, but, more important, the Gatti has shown
since the beginning a spatial dependence, especially along the radial direction. Figure~\ref{fig:gatti-r} shows, indeed, the shift of the $G_{\alpha,\beta}$ as a function of $r^3$ for \bipo events (N.B.: plotting data as a function of $r^3$ remove the spherical volume dependence over $r$). This dependence is neither easily modelled nor completely reproducible in Monte Carlo simulations. 
\begin{figure}[!h]
    \centering
    \includegraphics[width=1.0\columnwidth]{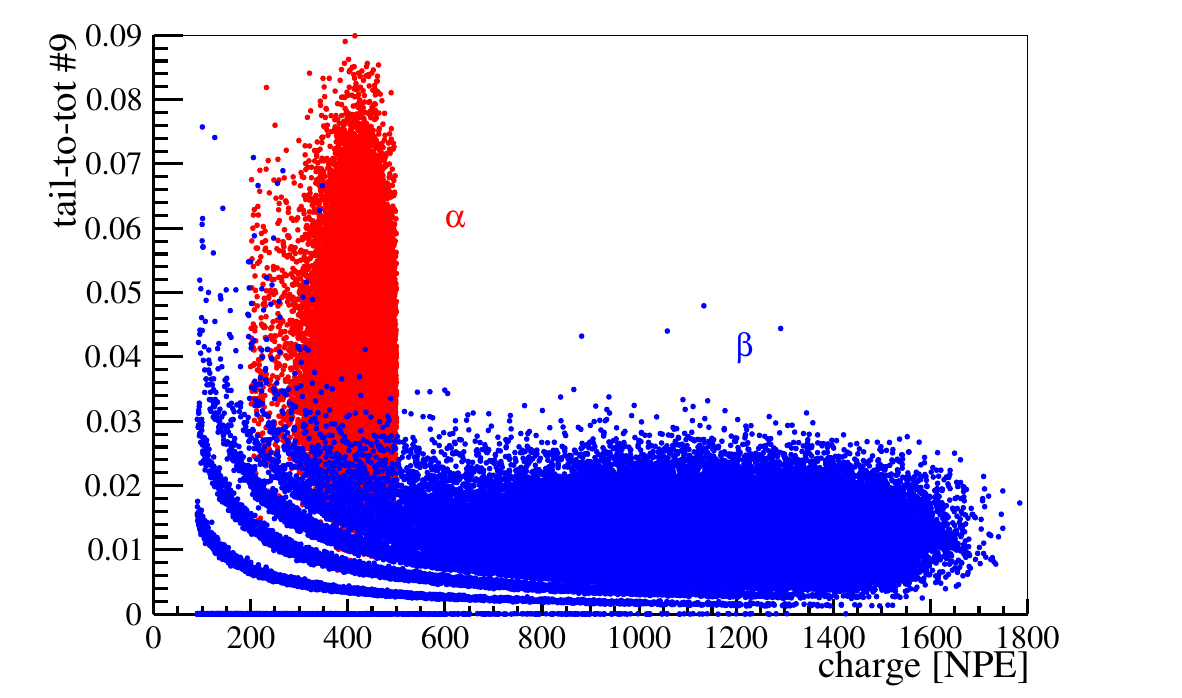}
    \includegraphics[width=1.0\columnwidth]{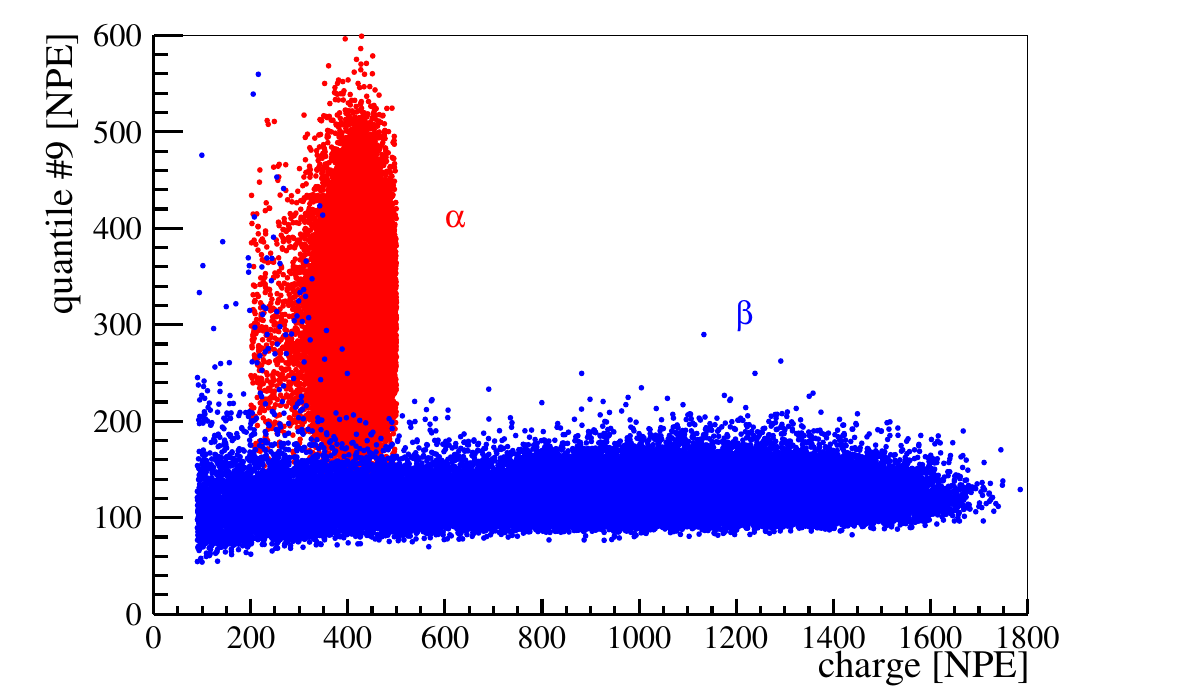}
    \caption{Limitations of the \texttt{t2t} input variable in comparison with quantiles. \emph{Top:} last \texttt{t2t} ($t > 310$ [ns]) as a function of the energy in NPE for $\alpha$'s (red) and $\beta$'s distribution. \emph{Bottom:} same distributions for the last quantile (10 \% tail of time PDF).}   
    \label{fig:tailtot_quantile_charge}
\end{figure}
\begin{figure*}[!t]
    \centering
    \includegraphics[width=1.0\textwidth]{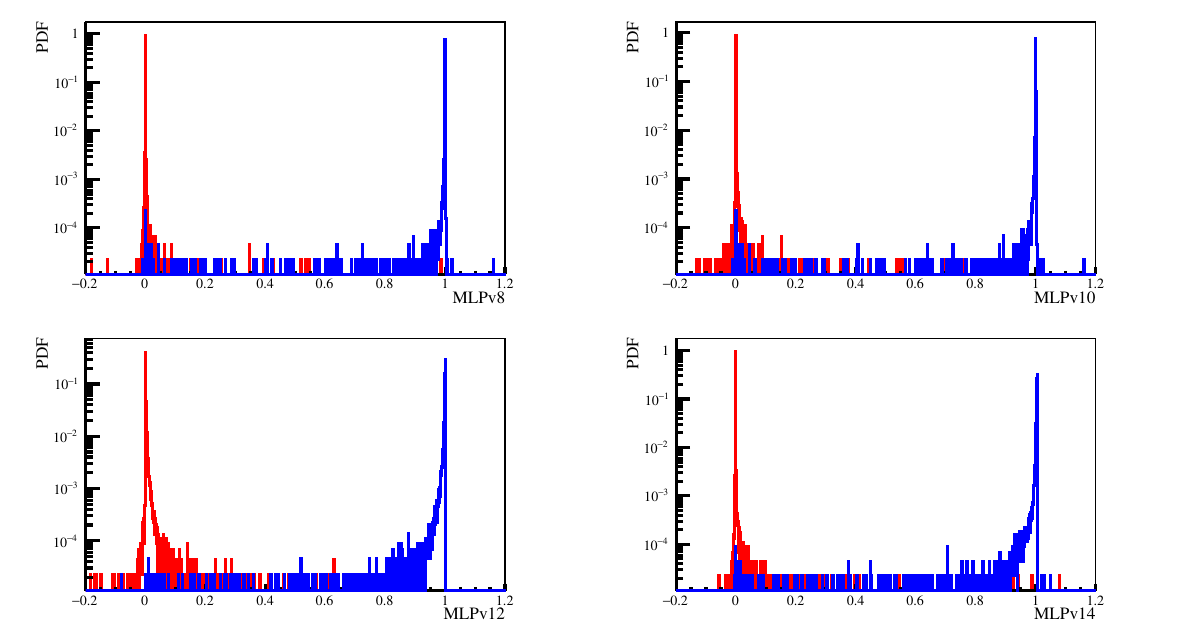}
    \caption{Comparison between MLP versions with test \texttt{Sample-WEX}. Form the top left: \texttt{MLPv8}, \texttt{MLPv10}, \texttt{MLPv12} and \texttt{MLPv14}. Red and blue PDFs represent $\alpha$'s and $\beta$'s events, respectively.}
    \label{fig:cmp}
\end{figure*}
Contrary to the Gatti filter, artificial neural networks, accepting plenty of input parameters, and returning their corresponding ranking for the specific case of the \ab selection, helped a lot in understanding the origin of the radial dependence of the PSD. They offered a more uniform selector, with a controllable dependence of the efficiency upon energy and time, as discussed later in Sec.~\ref{sec:3subref}. In the next Section, the strategy for the implementation and tuning of a class of \emph{multi-layer perceptron} is reviewed.

\section{Improving the selection with MLP}
\label{sec:mlp}

\subsection{Artificial neural networks}

\emph{Artificial neural networks} (ANNs) are a major component of machine learning and are designed to detect patterns in data~\cite{bib:ANN}. This makes ANNs the optimal solution for classifying (sorting data by predetermined categories), grouping (finding similar characteristics among data and combining that data into categories), and making forecasts based on the data.

An ANN is, more generally speaking, any simulated collection of interconnected neurons, with each neuron producing a certain response at a given set of input signals. The input data can be values for the characteristics of an external data sample, such as images or documents, or it can be outputs from other neurons. 

By applying an external signal to some input neurons, the network is put into a defined state that can be measured from the response of one or several output neurons. One can therefore view the neural network as a mapping from a space of input variables $x_1,...,x_{n_{\rm var}}$ onto a \emph{one-dimensional}
(\emph{e.g.} in case of a signal-versus-background discrimination problem) or \emph{multi-dimensional} space of output variables. The mapping is nonlinear if at least one neuron has a nonlinear response to its input.
It is important noticing here that Gatti parameter is linear over the input parameters, as one can easily see from Eq.~(\ref{eq:GattiDef}), therefore, given the input data set, it cannot do better than their intrinsic statistical power.

A multilayer perceptron (MLP) is a class of feed-forward artificial neural network.
An MLP consists of at least three layers of nodes: an input layer, a hidden layer and an output layer.  Except for the input nodes, each node is a neuron that uses a nonlinear activation function.  MLP utilises a supervised learning technique called \emph{back-propagation} for training.

\subsection{TMVA package}

The  Toolkit  for  Multivariate  Analysis  (TMVA)  provides  a  ROOT-integrated  environment  for the processing, parallel evaluation and application of multivariate classification techniques \cite{bib:tmva}. TMVA is specifically designed for the needs of high-energy physics (HEP) applications where the search for ever smaller signals in ever larger data sets has become essential to extract a maximum of the available information from the data. Multivariate classification methods based on machine learning techniques have become an essential ingredient in most of the HEP analyses. The package hosts a large variety of multivariate classification algorithms, \emph{e.g.} artificial neural networks (three different MLPs implementations), support vector machines (SVM), boosted decision trees (BDT), \emph{etc.}

Independent input data sets used for \emph{training} and \emph{testing} of multivariate methods must be defined prior to the algorithm implementation. The most important step is to identify such input parameters that are important in order to obtain the highest possible efficiency of the pulse shape discrimination of signals, and this can be done looking at the returned ranking of the variable themselves at each controlled trial.

\subsection{Selection of input variable and different versions}
\label{sec:3subref}

It is standard practice to normalise the input variables before integrating them into the ANN. In the Borexino case of \ab discrimination, a set of \texttt{t2t} variables were defined for ten different $t_0$, according to Eq. (\ref{eq:t2t}).
Due to the fact that the distributions for $\alpha$ and $\beta$ (Fig.\ref{fig:abPulse}) differ mainly in the tails, times after 10 ns were chosen, i.e. $t_0$ in the set $\{35, 70, 105, 140, 175, 210, 245, 280, 315,350\}$ (ns). To this set, the root mean square (RMS) and kurtosis of the photoelectron time distribution were added. 
At this stage, having a set made of \texttt{t2t}'s, RMS and kurtosis in the input vector, the MLP algorithm returns the discrimination efficiency similar to Gatti, as expected.
The statistical theory is absolutely constraining here, so it was necessary to try to add information not present in the first trial.
\begin{figure}[!h]
    \centering
    \includegraphics[width=1.0\columnwidth]{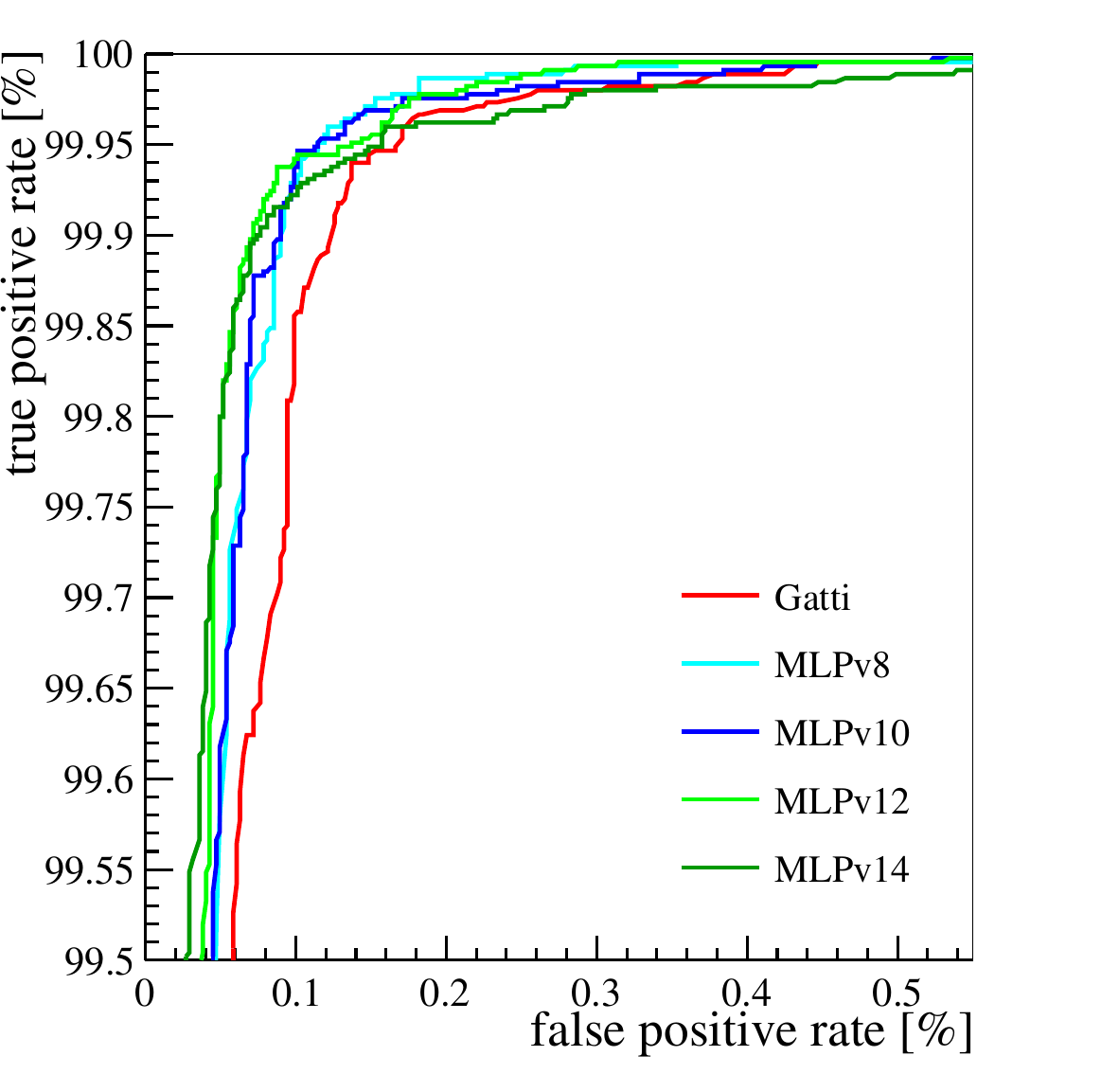}
    \caption{Zoom of the ROC curves around the region of interest for Gatti and four MLP versions according to the color legend inset. The analysis is performed over the \texttt{Sample-WEX} test sample.}
    \label{fig:roc}
\end{figure}

A breakthrough came when it was noticed that the time distribution of the scintillation events is analysed after the event position correction, determined by the time-of-flight of photons originated in a point-like scintillation and propagated towards the PMTs. From subsequent trials, it was observed that the \texttt{mean-time} variable of the hits calculated before the position reconstruction (``non reconstructed cluster'') was adding some missing information, possibly lost with the position correction. This recovered information improved the \ab discrimination, even solving the radial dependence observed in the Gatti parameter. This \texttt{mean-time} is basically the mean of the temporal PDF of the scintillation events, in which the times are associated with the photomultiplier reference system.

This finding also clarified why in CTF the \ab discrimination was working in a more efficient way. In practice, since the CTF detector was a few meters small, there was basically no bias due to off-centre event reconstruction. This guess is also confirmed for events located in a region very close to the centre of Borexino, where the Gatti parameter does not show a substantial bias, and exhibits a very high efficiency.

\begin{figure}[!h]
    \centering
    \includegraphics[width=1.0\columnwidth]{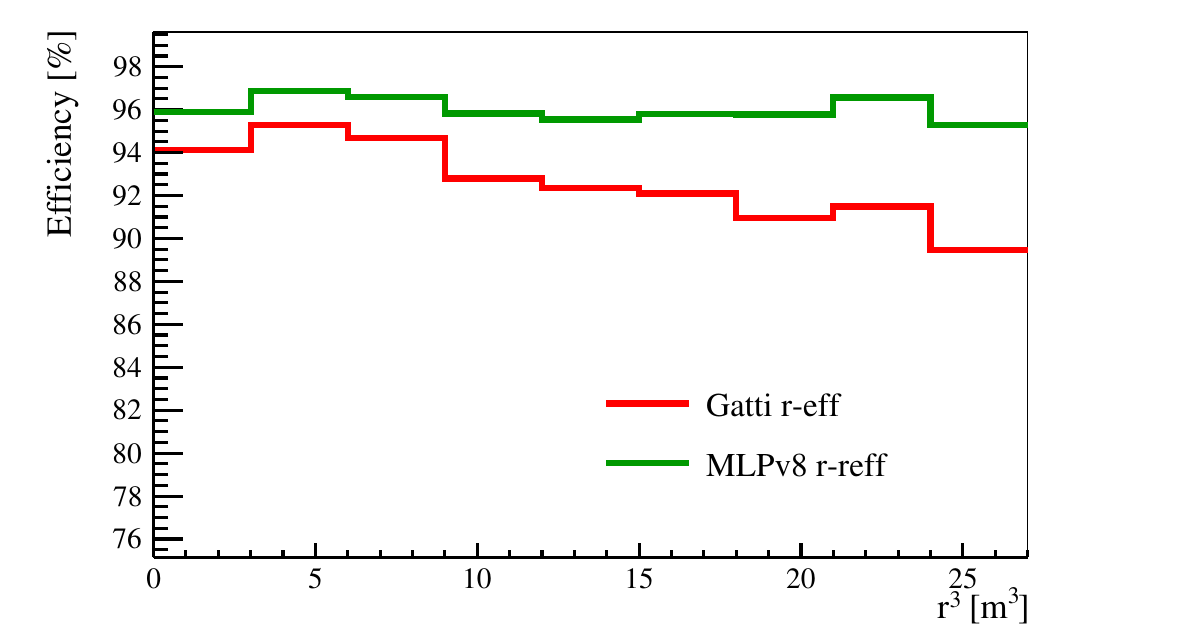}
    \caption{Comparison between the radial efficiencies of the PSD cut for  \texttt{MLPv8}$<0.3$ (green) and Gatti$>0$ (red). In order to account for the cubic increase of the statistics for the radial distribution, the efficiency is reported as a function of $r^3$, instead of $r$.}
    \label{fig:radial}
\end{figure}

\begin{figure}[!h]
    \centering
    \includegraphics[width=1.0\columnwidth]{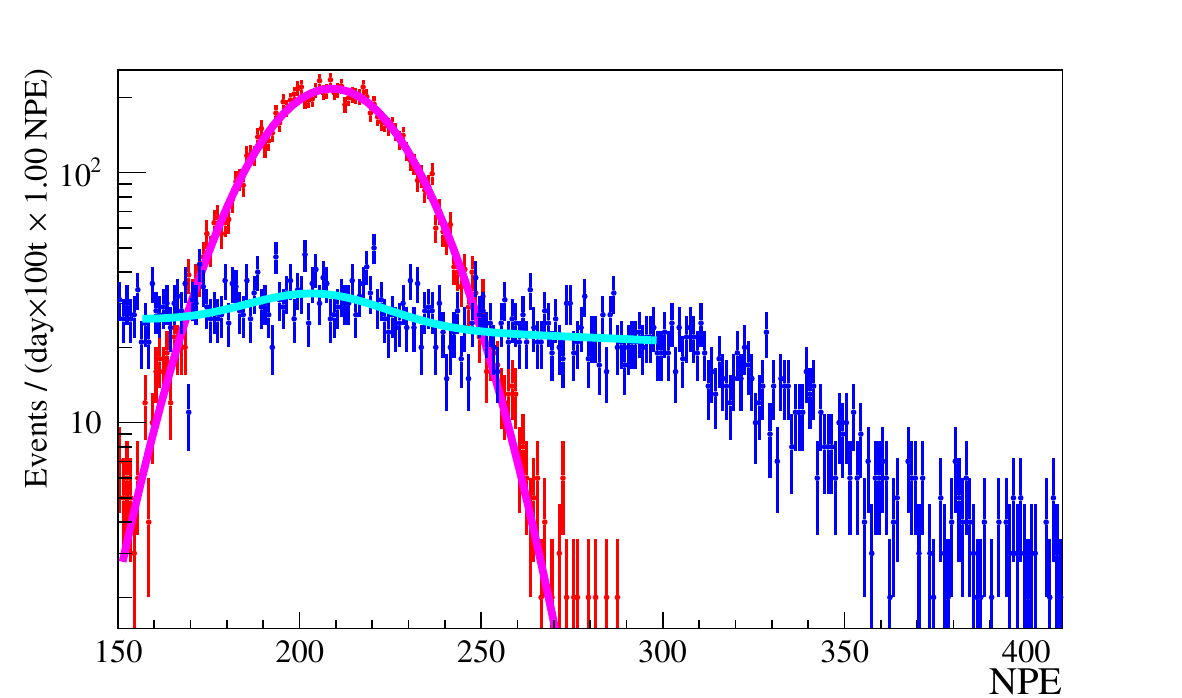}
    \caption{Example of `MLP-subtracted" (red) and ``MLP-complementary'' (blue) histogram fitted with the MLP-complementary method. The red and blue distributions represent $\alpha$ and $\beta$ complementary sets, whereas the violet and cyan curves represent the fit models accounting for the exponential dependence of the MLP cut as in Eq.~\ref{eq:ansatz}. This analysis is performed over one-year period of Phase-II, using \texttt{MLPv8}.}
    \label{fig:MLPcmp}
\end{figure}

A further improvement is achieved in MLPs in which the ten \texttt{t2t}'s input variables are replaced with the ten PDF quantiles. Quantiles gives same statistical weight to the input variables, being indeed defined as one tenth of PDF area. This definition avoids numerical quantisation problem of \texttt{t2t}'s, coming from the integer definitions of $t_0$'s (see Fig.~\ref{fig:tailtot_quantile_charge}) and, more important, remove the correlation present by definition on the \texttt{t2t} inputs, basically due to the partial overlap of the integrals for different $t_0$.

\subsection{MLP Test and Training from $^{222}$Rn events.}

The Gatti parameter in Borexino was initially tuned on 7000 \bipo events collected during the scintillator operations, before the official start of data acquisition in mid-2007. Subsequently, during six cycles of the WE campaign, occurred in between 2010 and  2011, another and bigger sample of \bipo events was collected. This sample, on which the Gatti parameter was upgraded and the MLP studies are based, contains in total 85.000 events, whose 27000 events lies in the fiducial volume region $r\simeq 3$ m. The only issue, that must be taken into account and controlled, is the fact that, in both data-sets (Phase-I and WE), the radon events were observed mainly on the detector top, in the region above the equator ($z>0$). This evidence, supported also by the fluid dynamic simulations performed for the CNO analysis~\cite{bib:cno}, is a consequence of the effective separation between the two hemispheres due to the fluid motion in relation with the spherical geometry. After the MLP training, a slight top-bottom asymmetry was actually observed, but was found not of practical relevance inside the analysis fiducial volume.

The final training sample (\texttt{Sample-WE}) contains in each MLP version 25000 
events with $r<3$ m from the WE period. The comparison of performances, among MLPs and Gatti, was done
on a reduced sample made of about 15000 events for training, and about 15000 events for test (\texttt{Sample-WEX}), both with larger radii to study also the radial dependence. Another test samples (\texttt{Sample-Ph23}), for double-checking the evolution of the efficiency in time, space and energy, was chosen selecting an \ab sample, not from the \bipo (basically absent after WE), but from two energy intervals of the Borexino Spectrum in the first 1000 days of Phase-II, in which the contribution of the \po activity is still sizeable. The $\alpha$ sample is selected in a very narrow region of the \po peak (209-210 NPE), with a very small contamination of the underlying $\beta$-like component from solar neutrino and $\beta$ decays; whereas the $\beta$ sample is selected in the \be shoulder region (320-400 NPE), where the leakage of the $\alpha$ events from the \po right tail is also negligible.

As anticipated, the MLP after the WE period, shows a performance degradation because of PMT loss with corresponding degradation of event reconstruction resolution, resulting in a time and space (radial) dependence. Furthermore, the $^{214}$Po emits a mono-energetic alpha line about 50\% higher than the \po peak energy, that falls indeed in the region of interest for the solar neutrino analysis. As a consequence of the energy dependence of the scintillation temporal PDFs, the MLP efficiency evaluated at the $^{214}$Po line is not directly applicable for example on the \po analysis. The correct assessment of the space, time, and energy dependence of the MLP was studied using also calibration data and Monte Carlo simulations. In the following paragraph we will report the main MLP features investigated in the Borexino analysis.  

\subsection{Different versions of \ab MLP}

The most important MLP versions, which gave similar performances and were used in the main Borexino analyses, are listed here:

\begin{enumerate}
    \item \texttt{MLPv8}: This version is the first showing a significant improvement with respect to Gatti. The input variable are the ten \texttt{t2t}'s described above, in addition with the RMS, the kurtosis and \texttt{mean-time} of the non-reconstructed cluster.  
    \item \texttt{MLPv10}: This version is similar to \texttt{MLPv8}, but \texttt{t2t}'s are replaced with 10 quantiles. In some cases, this version shows a slightly better performance as compared with \texttt{MLPv8}, for the reasons described above, especially for low energy events.
    \item \texttt{MLPv12}: This version was meant to solve problems coming from the energy difference between the training \bi sample, and the low energy region where the \ab is actually applied. The \bi sample energy is artificially reduced by thinning out the number of photoelectron (randomly removed), in a ratio of 1:2 for the $^{214}$Po and of 1:4 for $^{214}$Bi. Although this method is assigning the correct statistical weight to the training samples (and so similar to the low energy region), it cannot include real energy dependence of the scintillation PDFs upon the energy. 
    \item \texttt{MLPv14}: Finally this version, attempts to solve the problem of the low energy extrapolation using $^{218}$Po events for $\alpha$'s (with lower energy and then closer to \po). Since the $^{218}$Po precedes the \bipo fast coincidence by about 30 minutes, a space time cut (1 meter radius and 1.5 hour before) was able to select a pure sample of about 1800 event candidates. Due to the very low efficiency of the $^{218}$Po tagging, this sample contains on one hand a proper representative set of low energy events, but on the other hand it has a limited statistics. 
\end{enumerate}

In order to compare and contrast the different versions of MLPs, several studies have been performed, especially in terms of efficiency, space uniformity and time stability.

The TMVA package returns the normalised selector in the 0--1 interval, sharply peaked at 0 for $\alpha$'s and at 1 for $\beta$'s in the Borexino choice.
Figure \ref{fig:cmp} shows the distribution of the 0--1 selector for the versions of interest for $\alpha$'s (red) and $\beta$'s (blue) from test \texttt{Sample-WEX}. All of them are basically comparable, even if \texttt{MLPv8} shows in general a better symmetry and sharper distributions. \texttt{MLPv12}, for the reasons discussed above, shows a more smeared distribution, even though with a good separation. This differences can be understood comparing the same discriminator with \emph{Receiver Operating Characteristic} (ROC) plot as reported in Fig. \ref{fig:roc}. In particular, assuming $\alpha$'s as signal and $\beta$'s as background (N.B.: opposite to the typical TMVA convention in this particular analysis), the ROC curve reports the \emph{True Positive Rate} (TPR) as function of the \emph{False Positive Rate} (FPR) changing the selection threshold $m_0$ in the interval $0 < m_0 <1$, that is
\begin{eqnarray}
{\rm TPR} & = & \int_0^{m_0} \mathcal{M}_\alpha (t) dt, \\
{\rm FPR} & = & \int_0^{m_0} \mathcal{M}_\beta (t) dt, 
\end{eqnarray}
where $\mathcal{M}_{\alpha, \beta}$ are the corresponding PDFs of the MLP parameters, always from the test \texttt{Sample-WEX}. From Fig. \ref{fig:roc}, one can compare the overall performance of different MLP versions (bluish and greenish curves) and the Gatti parameter (red), as reported in the Figure legend. The better discriminator approaches a right angle shape in the top-left corner. From this Figure, we can conclude that all MLP have an overall good performance, considerably better than Gatti for FRP $< 0.2\%$, \emph{e.g.} for 99.75\% TPR one can have a factor 2 less contamination from FPR.

\subsection{MLP radial dependency}

The radial dependence of the MLP selector, strictly related to the position dependence of the reconstructed cluster, plays a crucial role in the \po spatial analysis, as described at the end of Sec.~\ref{sec:gatti}. It is therefore important to study, through Monte Carlo simulations and through the test samples, any possible feature of the MLP related to the position and its possible bias in the \po activity determination. Figure~\ref{fig:radial} shows the radial efficiency determined from the test \texttt{Sample-Ph23} for \texttt{MLPv8} (green) and Gatti (red): the first shows a better behaviour in terms of spatial uniformity. If one consider the CNO fiducial volume, located at about 21 $m^3$ on the $x-$axis, the efficiency is pretty uniform. As discussed in~\cite{bib:cno}, the non-uniformity of the \texttt{MLPv8} efficiency is indeed negligible, as compared with the energy and time dependence, which will be discussed below.

Such studies have been performed with different methods and with different values of the MLP selection threshold.

\begin{figure}[!h]
    \centering
    \includegraphics[width=1.0\columnwidth]{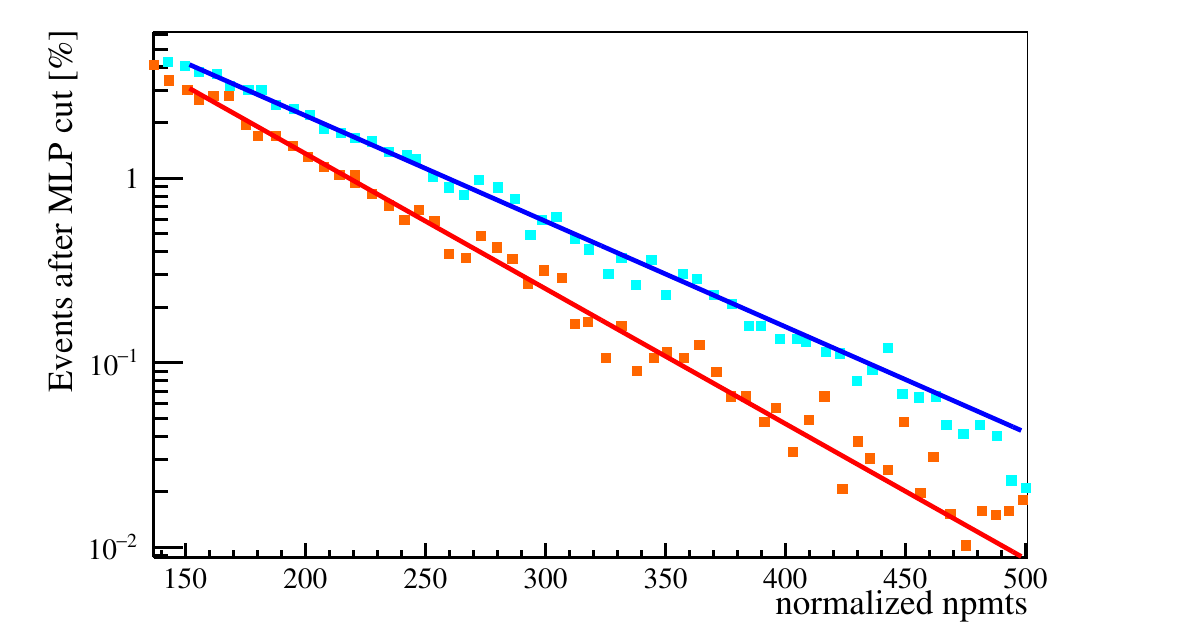}
    \caption{Example of energy dependence (\texttt{npmt} energy estimator) of the MLP efficiency, performed over Monte Carlo data. The $\beta$-like event percentage left after the MLP$<0.015$ cut are reported for the versions \texttt{MLPv12} (orange) and \texttt{MLPv10} (cyan). Both curves are fitted to exponential functions, namely the red and the blue lines in the logarithmic plot.}
    \label{fig:eff1}
\end{figure}

\subsection{Stability and energy dependence of MLPs}

The time dependency of the MLP, for a given selection cut, in the \po energy region has been carefully studied for the time stability of the \bi and \po activity in the context of the CNO neutrino analysis, see Sec.~\ref{sec:cno}. In order to obtain the selection efficiency of \po and the corresponding leakage of $\beta$ events by the cut itself, events in the fiducial volume analysis are fitted with the so-called \emph{MLP-complementary} method. In the latter, the data-set, year by year from 2011, are split into two histograms depending on events passed or not the MLP cut, named ``MLP-subtracted" and ``MLP-complementary'', respectively, as reported in Fig. \ref{fig:MLPcmp}.  

\begin{figure}[!h]
    \centering
    \includegraphics[width=1.0\columnwidth]{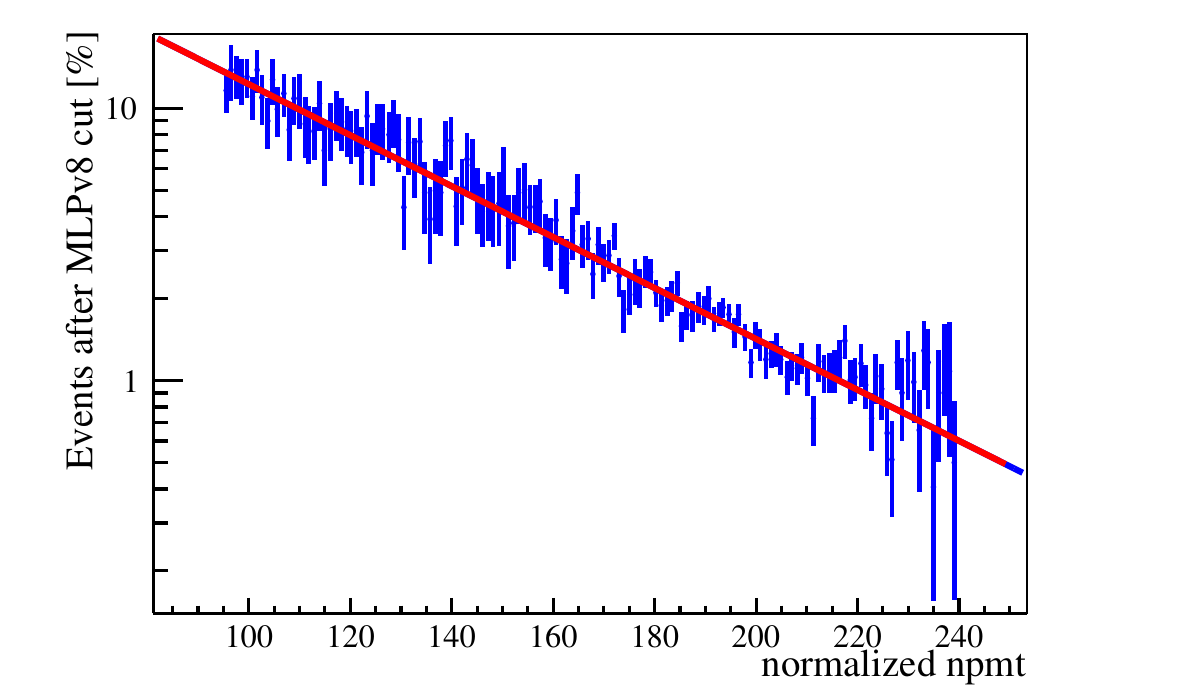}
    \caption{Energy dependence  (\texttt{npmt} estimator \cite{bib:long}) of the fraction of $\beta$ events left after the \texttt{MLPv10} cut on real $\beta$-like calibration data.}
    \label{fig:eff2}
\end{figure}

For typical analysis, a PSD threshold of $m_0<0.05$ and energy in 150-300 NPE interval are used. Under these conditions, the fitted spectra, as a function of the energy $E$ [NPE], are defined as
\begin{eqnarray} \label{eq:ansatz}
S_\beta(E) & = & S_{\rm bx}(E) (1 - A e^{-E/E_0}), \\
S_\alpha(E) & = & S_{\rm bx} (E) (A e^{-E/E_0}),
\end{eqnarray}
where $S_{\rm bx}(E)$ is the typical Borexino spectrum with all fitted species (see \emph{e.g.} \cite{bib:nusol}), $S_{\alpha, \beta}$ are the resulting $\alpha$ and $\beta$ selected spectra and, finally,  $A$ and $E_0$ are two free parameters. Notice that the ansatz that the energy dependence of the MLP cut is exponential is suggested by Monte Carlo simulations and calibration data, and also by general considerations about the statistical nature of the neural network output. Figures~\ref{fig:eff1} and \ref{fig:eff2} show the exponential energy dependence of the MLP cut from Monte Carlo simulation and from calibration data, respectively. In both Figures, the energy estimator is \texttt{npmt}, i.e. the number of hit PMTs during a scintillation event without double counting piled-up events, see~\cite{bib:long} for further details.
Either in calibration data and in MC events, the percentages left after the MLP cut show an exponential behaviour, supporting the choice of the energy dependence of the efficiency assumed in the MLP-complementary fit.

\begin{figure}[!h]
    \centering
    \includegraphics[width=1.0\columnwidth]{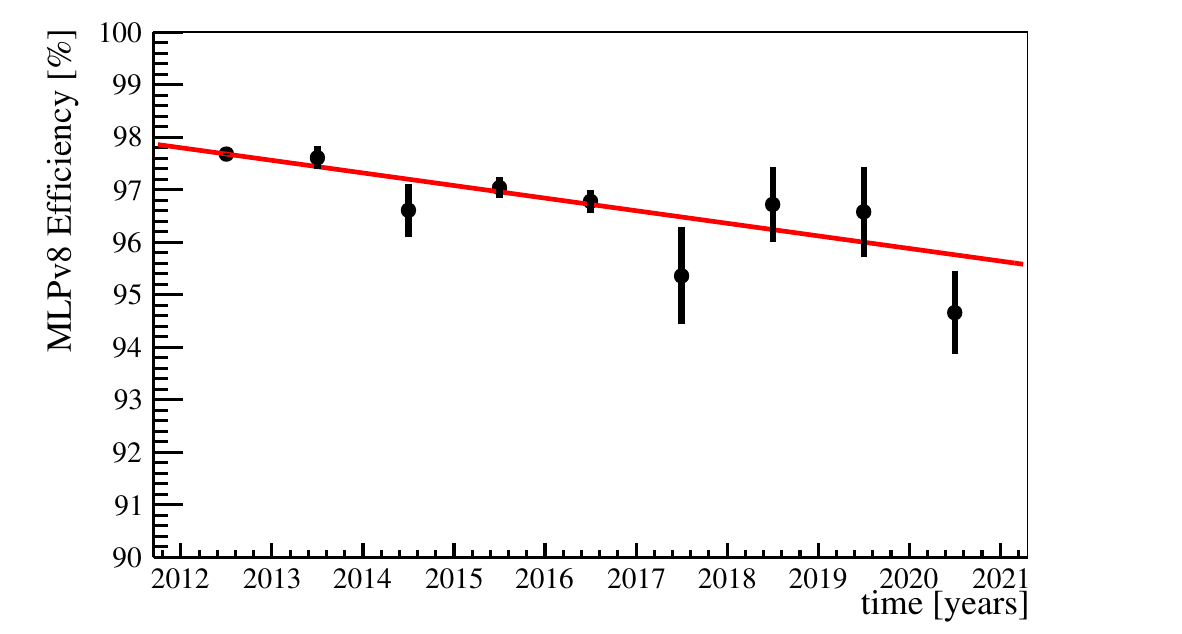}
    \caption{MLP time dependence efficiency from MLP-complementary fit for the combined Phase-II and Phase-III period for \texttt{MLPv8}, performed over one-year time intervals (blue histogram). A fitted linear trend (red) shows, at leading order, the degradation of the MLP efficiency due to the PMTs loss ($p$-value$=0.48$ ). The uncertainty increases over time because the reduced \po statistics due to its decay.}
    \label{fig:MLPtime}
\end{figure}

Finally, Fig.~\ref{fig:MLPtime} shows the time evolution of the MLP efficiency for the \po energy range year by year, resulting from the model in Eq. (\ref{eq:ansatz}). The slightly decreasing linear trend is compatible with expectation of the event reconstruction degradation, mainly related to the linear  PMT loss. This dependence is used to correct the measurement of the \po activity and to determine the corresponding systematic uncertainty for the final result on the CNO neutrino interaction rate.

\section{Event-by-event MLP tagging in the main Borexino analyses.}
\label{sec:cno}

\subsection{Polonium-210 studies for the CNO quest}

The possibility of tagging $\alpha$ events with high efficiency in space and time was of crucial importance for the first measurement of neutrinos from the CNO cycle~\cite{bib:cno} with Borexino and its subsequent update~\cite{bib:PRLcno}.

Given the degeneracy, and then the correlation, between \bi, \pep and CNO spectra, the sensitivity to CNO neutrinos through the spectral analysis is pretty poor, unless the \bi and $pep$-$\nu$ rates are independently constrained in the spectral fit\,\cite{bib:sensitivity-paper}.
In particular, the $pep$-$\nu$ rate can be constrained to 1.4\% precision~\cite{bib:sensitivity-paper}, using: solar luminosity along with robust assumptions on the $pp$ to $pep$ neutrino rate ratio, global analysis of existing solar neutrino data~\cite{bib:Vissani2019, bib:bergstrom}, and the most recent oscillation parameters\,\cite{bib:NuPars}.
The \pep constraint is essentially independent of any reasonable assumption on the CNO rate, as the solar luminosity depends only weakly on the contribution of the CNO cycle itself.

In practice, the only crucial element at play is the \bi rate, a $\beta$ emitter with a short half-life (5\,days) coming from the $^{210}$Pb (present in the scintillator at the beginning of Phase-II) through the decay chain:
\begin{equation} \label{chain210}
^{210}\mathrm{Pb} \xrightarrow[22.3\, \mathrm{years}]{\beta^-}{^{210}}\mathrm{Bi} \xrightarrow[5\,\mathrm {days}]{\beta^-}{^{210}}\mathrm{Po} \xrightarrow[138.4\,\mathrm{days}]{\alpha} {^{206}}\mathrm{Pb}\,.
\end{equation}
Assuming the secular equilibrium, the \bi rate can be determined from the \po activity
\cite{bib:secular, Villante2011, bib:sensitivity-paper}. Since the \po activity can be measured precisely through the MLP high efficiency \ab tagging, this strategy provided the key solution to tackle the species correlation in the spectral fit and then to lead the Borexino Collaboration to the first observation of the CNO neutrino interaction rate and its subsequent upgrade~\cite{bib:cno, bib:PRLcno}. Anyway, the story was not that easy: at the beginning of Borexino Phase-II (early 2013), it was clear that presence of convection motions, caused by the seasonal change of the temperature in the Gran Sasso experimental hall, made it impossible to apply the \bi-\po link, as suggested by the sequence in Eq.~(\ref{chain210}). 

In order to solve this problem, a long and challenging thermal stabilisation program was undertaken by the Collaboration to prevent the scintillation convctive motion, and the consequent contaminant mixing in the scintillator. This program, started in mid-2014, consisted of different phases: (i) installation of high precision temperature probes inside and outside the detector, (ii) thermal insulation of the detector with different layers of rock wool, (iii) active temperature control systems of the detector and (iv) of the experimental room. This longstanding effort worked properly and allowed one to set an upper limit on the \bi rate, a crucial ingredient for the final extraction of the CNO neutrino interaction rate from the spectral analysis.

It is worth mentioning that the MLP tagging, with its high efficiency, uniformity and stability, helped in all the stages of this enterprise: from the understating of the \po migration in the scintillator, through study of the effects of the different phases of the thermal insulation program, until the determination of the \bi upper limit rate (see for details the appendix of~\cite{bib:cno}).

For the CNO analysis the space and time dependence of the \ab tag in the \po region was studied carefully using the MLP complementary analysis and Monte Carlo simulations. In particular, the latter was crucial for the optimisation of the cut and for the efficiency dependency upon the radial position and time.
In particular the best cut was defined by maximising the standard signal-to-background (S/B) figure of merit (FoM): 
\begin{equation}
    {\rm FoM} = \frac{S}{\sqrt{S+B}}.
\end{equation}
In this case $S$ can be assumed as true positive events (real $\alpha$'s), and $B$ as false positive ($\beta$-like events leaking out from the distribution tail).
For \texttt{MLPv8} the best $\alpha$ cut, corresponding to the Phase-III data set,  was found at $m_0<0.3$.

\subsection{MLP in other analyses}

Besides the CNO analysis, the MLP \ab tagging was used in many other analyses published by the Borexino Collaboration.  In particular, it played an important role in the high significance detection of the seasonal modulation of $^{7}$Be neutrinos due to the Earth orbit eccentricity \cite{bib:seas}, for the reduction of the \po component in the region of interest of the \be spectrum. This analysis was updated including the entire Phase-II and Phase-III data set, leading to the first independent measurement of the Earth orbit eccentricity with only solar neutrinos~\cite{bib:ecce}.

In the geo-neutrino analysis, the MLP was used in the event selection with a high performance even at large radii ($\sim$4m) close to the Nylon vessel \cite{bib:geonu} and for the \po background estimation for the neutron background induced by alpha decays. In addition, the MLP selection was used for the space and time selection of the \po data events for the accurate tuning of Monte Carlo used for simulating the \po spectrum in Phase-II~\cite{bib:mc} and Phase-III. In particular, this study played an important role in the comprehensive analysis of the \pp chain \cite{bib:pp, bib:nusol}. 
This study has provided a measurement of the most important solar neutrino fluxes, which is in 
favour of the MSW-LMA neutrino oscillation scenario at 98\% CL. (see \emph{e.g.} \cite{bib:msw} and Refs. therein).

\section*{Conclusions}

In this paper, we offer a detailed review of the \ab pulse shape discrimination adopted in Borexino. We present various implementations used during more than a decade of data taking, starting with the Gatti optimal filter and the corresponding statistical subtraction of the $\alpha$ component from the energy spectrum, ending with the more sophisticated PSDs based on ANNs, specifically exploiting MLP. The latter, with its high efficiency, spatial uniformity and time stability, allowed us to event-by-event select the \po events, a crucially important background reduction which made possible the observation of CNO solar neutrinos in Borexino. 

Compared to the Gatti parameter approach, ANNs single out parameters relevant to PSD in a highly non factorizable way.
In the case of Borexino, the Gatti parameter was limited by information loss in the photon arrival times after position reconstruction correction.
%The improvement with respect to the Gatti parameter is ascribable essentially to the capability of the ANN to single out parameters that play the most crucial role in PSD. 
%In the case of Borexino, the Gatti parameter was limited by information loss due to photoelectron arrival times corrected with the position reconstruction. 
The integration of variables in the MLP before event reconstruction improved the performance of the \ab selection.
The MLP implementation required careful calibration to select the best input parameters, tune the algorithm, and evaluate its performance.
%dedicated studies, especially for the selection of the input parameters, the tuning and the assessment of the performances. 
Its spatial and time efficiency were monitored and used for the evaluation of the global systematic uncertainty of some of the most important Borexino results for which the method was used.

The \ab pulse shape discrimination allowed by intrinsic properties of the 
%Borexino 
scintillator, was proven to be fully exploitable in an ultra-pure, large-volume detector such as Borexino. In particular, it played an essential role in the neutrino spectroscopy of the entire \pp chain and the first observation of neutrinos from the CNO cycle.

\section*{Acknowledgements}

We acknowledge the generous
hospitality and support of the Laboratori Nazionali del Gran
Sasso (Italy). The Borexino program is made possible by
funding from Istituto Nazionale di Fisica Nucleare (INFN)
(Italy), National Science Foundation (NSF) (USA), Deutsche
Forschungsgemeinschaft (DFG), Cluster of Excellence
PRISMA+ (Project ID 39083149), and recruitment initiative
of Helmholtz-Gemeinschaft (HGF) (Germany), Russian
Foundation for Basic Research (RFBR) (Grants No.
19-02-00097A), Russian Science Foundation (RSF) (Grant
No. 21-12-00063) and Ministry of Science and
Higher Education of the Russian Federation (Project
FSWU-2023-0073) (Russia), and Narodowe Centrum Nauki
(NCN) (Grant No. UMO-2013/10/E/ST2/00180) (Poland).
We gratefully acknowledge the computing services of
Bologna INFN-CNAF data centre and U-Lite Computing
Center and Network Service at LNGS (Italy). This 
research was supported in part by PLGrid Infrastructure (Poland).

\end{document}